\documentclass[12pt]{article}
\usepackage{amssymb}
\usepackage{color,graphicx}
\usepackage{amsmath}
\usepackage{axodraw}

\newcommand{\ep}{\varepsilon}
\newcommand{\bea}{\begin{eqnarray}}
\newcommand{\eea}{\end{eqnarray}}
\newcommand{\be}{\begin{equation}}
\newcommand{\ee}{\end{equation}}

\begin{document}

\begin{center}
{\Large {\bf The property of maximal transcendentality:
calculation of Feynman
integrals
}}
\\ \vspace*{5mm} A.~V.~Kotikov
\end{center}

\begin{center}
Bogoliubov Laboratory of Theoretical Physics \\
Joint Institute for Nuclear Research\\
141980 Dubna, Russia
\end{center}

\begin{abstract}
We review 
some results of calculations,
having the property of maximal transcendentality.

\end{abstract}


\section{Introduction}

It is well known 
that the 
popular  property of maximal transcendentality, which was introduced in \cite{KL}
for the Balitsky-Fadin-Kuraev-Lipatov (BFKL) kernel \cite{BFKL,next} in the  ${\mathcal N}=4$
Supersymmetric Yang-Mills (SYM) model \cite{BSSGSO}, is also applicable for the anomalous dimension (AD) matrices
of the twist-2 and twist-3 Wilson operators and for the coefficient functions of the ``deep-inelastic scattering'' (DIS)
in this model.
The property gives a possibility to recover the results for the ADs \cite{KL,KoLiVe,KLOV} and  the coefficient functions
\cite{Bianchi:2013sta}
without any direct calculations by using the QCD corresponding values
\cite{VMV}.

The
very similar
property appears
also in the results of calculation of the large class of Feynman integrals (FIs), mostly 
for  so-called  master integrals \cite{Broadhurst:1987ei}.
The results for most of them can be reconstructed also without any direct calculations using a knowledge of several terms
in their inverse-mass expansion \cite{FleKoVe}. Note that the properties of the results are related with the ones of
the amplitudes, form-factors and correlation functions (see \cite{Eden:2011we,Schlotterer:2012ny,Eden:2012rr} and
references therein) studied recently  in the framework of the  ${\mathcal N}=4$ SYM.

In this brief review, we demonstrate the existence of the propertiy of maximal transcendentality (or maximal complexity)
in  the
results of two-loop two- and three-point FIs (see also \cite{Kotikov:2010gf}). Moreover, we show its manifestation
for the eigenvalues of AD  
matrices of the twist-2 Wilson operators.

\section{Calculation of Feynman integrals}

\begin {figure} [htbp]
\centerline{
\begin{picture}(150,100)(0,0)
\SetWidth{0.7}
\CArc(75,50)(30,0,180)
\Line(45,50)(25,50)
\Line(105,50)(125,50)
\CArc(75,50)(30,180,360)
\Vertex(75,90){3}
\Vertex(75,50){3}
\Vertex(75,10){3}
\SetWidth{2.0}
\Line(75,10)(75,90)
\end{picture}
%
~~~~~~~~~~~
\hspace{-3.5cm}
\begin{picture}(150,100)(0,0)
\SetWidth{0.7}
\CArc(75,50)(30,0,180)
\Line(45,50)(25,50)
\Line(105,50)(125,50)
\CArc(75,50)(30,180,360)
\Line(45,50)(105,50)
\Vertex(75,65){3}
\Vertex(75,90){3}
\Vertex(75,35){3}
\Vertex(75,10){3}
\SetWidth{2.0}
\Line(75,10)(75,90)
\end{picture}
~~~~~~~~~~~
\hspace{-3.5cm}
\begin{picture}(150,100)(0,0)
\SetWidth{0.7}
\CArc(75,50)(30,0,180)
\Line(45,50)(25,50)
\Line(105,50)(125,50)
\CArc(75,50)(30,180,360)
\Line(75,80)(75,20)
\Vertex(90,50){3}
\Vertex(75,90){3}
\Vertex(60,50){3}
\Vertex(75,10){3}
\SetWidth{2.0}
\CArc(125,50)(65,140,220)
\CArc(25,50)(65,320,40)
\Line(60,50)(90,50)
\end{picture}
}
\caption{The examples of usual and dual FIs.}
\label{sunsetMMm}
\end{figure}
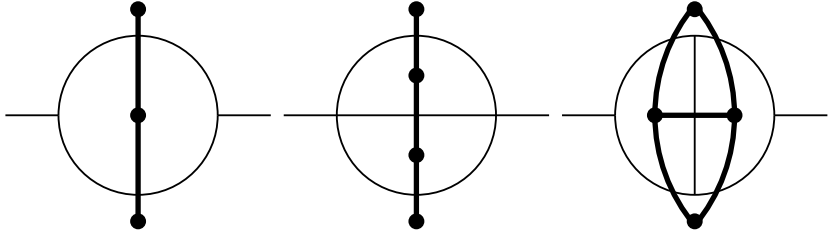

The arguments based on  the propertiy of maximal transcendentality
give a possibility to calculate
a large class of FIs in a simplest way.

Let us consider the results in some details.\\

{\bf 1.} At the beginning, we note that hereafter we will consider our FIs in the momentum space but it is
rather convenient also to work
in the {\it dual} coordinate space (see, for example, \cite{Kazakov:1986mu,Kotikov:1987mw,Kotikov:1995cw}), where
all the moments of the diagrams are replaced by the corresponding coordinates. Of course, the results of the integration
of the diagrams do not changed during the procedure.  However
the graphic representations 
are different. Shortly speaking, all loops (triangles, $n$-leg one-loop internal graphs)
should be replaced by the corresponding chains (three-leg vertices, $n$-leg vertices). For some simplest cases,
the replacement is shown on Fig. 1, where the thin lines correspond to the standard FIs
and the thick
ones show the corresponding dual graphs. More complicated cases were considered, for example, in
Ref. \cite{Kazakov:1984bw}. The integration in dual graphs are doing on the internal points. The rules of the
integration, including  the integration-by-part (IBP) procedure \cite{Chetyrkin:1981qh}, were considered
in  \cite{Kazakov:1986mu,Kotikov:1987mw}.

With the usage of the dual technique, the evaluation of the $\alpha_s$-corrections to the longitudinal DIS structure
function
has been done in \cite{Kazakov:1986mu,KK92}. All the calculations were done
for the massless diagrams. The extension of such calculations to the massive case were done in  \cite{Kotikov:1990zs}.
Some recent evaluations of the massive dual FIs can be found also in \cite{Henn:2014yza}.\\

{\bf 2.} Now we will return to the momentum space.
Application of the IBP procedure \cite{Chetyrkin:1981qh}
to loop internal momenta
leads to
relations between different FIs
and, thus,
to necessity to calculate only some of them, which in a sense, are
independent.
These independent diagrams
(which were chosen quite arbitrary, of course) are called the
master-integrals \cite{Broadhurst:1987ei}.

The application of the IBP procedure \cite{Chetyrkin:1981qh} to the 
master-integrals themselves leads to the differential equations (DEs)
\cite{Kotikov:1990zs,Kotikov:1990kg} for them with
the inhomogeneous terms (ITs) containing less complicated diagrams.
\footnote{The ``less complicated diagrams'' contain usually less number of
propagators and sometimes they can be represented as diagrams with less
number of loops and with some ``effective masses'' (see, for example, 
\cite{FleKoVe,Kniehl:2005bc} and references therein).} The application
of the IBP procedure to the diagrams in ITs leads to the new DEs
for them with the new ITs
containing even farther
less complicated diagrams ($\equiv$ less$^2$ complicated ones).
Repeating the procedure several times, at a last step one can obtain the
ITs containing mostly
tadpoles which can be calculated  in-turn very easily (see also the discussions in the part {\bf 3} below).

Solving the DEs
at this last step,
one can reproduce the diagrams for ITs of the DEs
at the previous step.
Repeating the procedure several times one can obtain the results for the
initial FIs.

This scheme has been used successfully for calculation of two-loop two-point
\cite{Kotikov:1990zs,Kotikov:1990kg,Fleischer:1999hp}
and three-point diagrams \cite{Fleischer:1997bw,FleKoVe}
with one nonzero mass. This procedure is very
powerful but quite complicated. There are, however, some simplifications, which
are based on the series representations of FIs.

\begin{figure}[t]
\includegraphics[width=0.6\textwidth,height=0.3\textheight,angle=0]{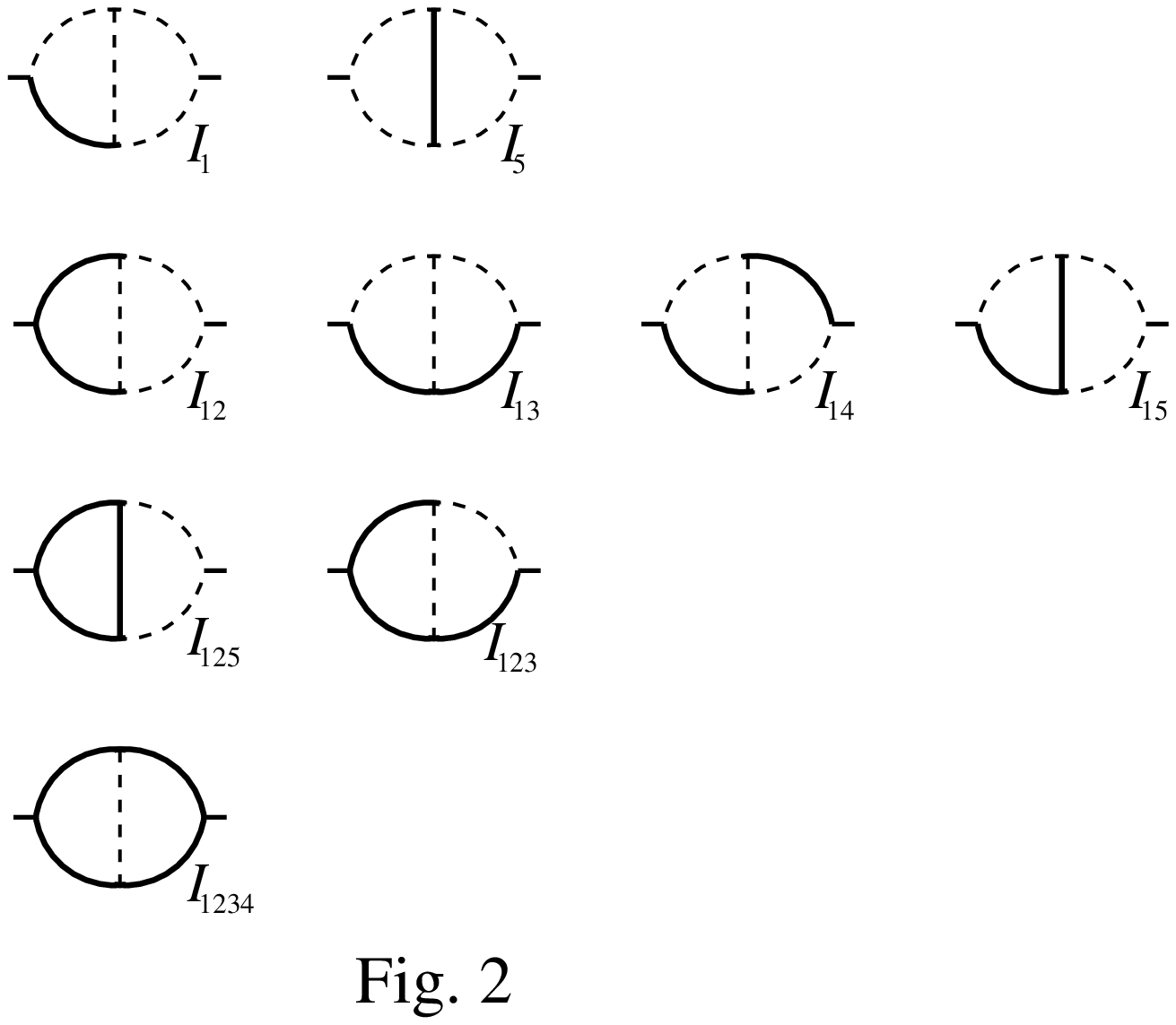}
\end{figure}

Indeed, the inverse-mass expansion of two-loop two-point (see Fig. 2)
and three-point
diagrams (see Fig. 3)
\footnote{The 
diagrams shown in Figs. 2 and 3, are 
complicated
two-loop FIs,
which have no three-massive-particle cuts. So, 
their results should be expressed as combinations of Polylogarithms. 
Note that we
consider only three-point
diagrams with independent upward momenta $q_1$ and $q_2$, which obey
the conditions
 $q_1^2=q_2^2=0$ and $(q_1+q_2)^2\equiv q^2 \neq 0$, where $q$ is downward
momentum.}
with one nonzero mass (massless and massive propagators are shown as dashed and solid lines,
respectively),
can be considered as
\begin{eqnarray}
&&\mbox{ FI} ~ 
= ~ \frac{\hat{N}}{q^{2\alpha}} \,
\sum_{n=1} \, C_n \, {(\eta x)}^n
\, \biggl\{F_0(n) +
\biggl[ \ln (-x) \, F_{1,1}(n)  +
\frac{1}{\varepsilon} \, F_{1,2}(n) \biggr] 
\label{FI1} \\
&& + \biggl[ \ln^2 (-x) \, F_{2,1}(n)  + \frac{1}{\varepsilon} \,\ln (-x) \,
 F_{2,2}(n) + \frac{1}{\varepsilon^2} \, F_{2,3}(n) + \zeta(2)
 \, F_{2,4}(n) \biggr]
+ \cdots \biggr\},
\nonumber
\end{eqnarray}
where $x=q^2/m^2$, $\eta =1$
or $-1$
and $\alpha=1$ and $2$ for
two-point and three-point cases, respectively.

Here the normalization 
$\hat{N}={(\overline{\mu}^2/m^{2})}^{2\varepsilon}$, 
where $\overline{\mu}=4\pi e^{-\gamma_E} \mu$ is in the standard
$\overline{MS}$-scheme and $\gamma_E$ is the Euler constant.
Moreover, the space-time dimension is $D=4-2\varepsilon$ and
\begin{eqnarray}
C_n  ~=~ \frac{(n!)^2}{(2n)!} ~\equiv ~ \hat{C}_n
\label{FI1b}
\end{eqnarray}
for diagrams with
two-massive-particle-cuts ($2m$-cuts). For the diagrams with one-massive-particle-cuts ($m$-cuts)
$C_n = 1$.

For  $m$-cut
case,
the coefficients $F_{N,k}(n)$ should have the form
\begin{eqnarray}
F_{N,k}(n) ~ \sim ~ \frac{S_{\pm a,...}}{n^b}\, ,  \frac{\zeta(\pm a)}{n^b}\, , 
\label{FI1c}
\end{eqnarray}
where $S_{\pm a,...} \equiv S_{\pm a,...}(j-1)$
are harmonic sums
\be
S_{\pm a}(j)\ =\ \sum^j_{m=1} \frac{(\pm 1)^m}{m^a},
\ \ S_{\pm a,\pm b,\pm c,\cdots}(j)~=~ \sum^j_{m=1}
\frac{(\pm 1)^m}{m^a}\, S_{\pm b,\pm c,\cdots}(m),  \label{FI2}
\ee
and $\zeta(\pm a)$ are the Euler-Zagier constants
\be
\zeta(\pm a)\ =\ \sum^{\infty}_{m=1} \frac{(\pm 1)^m}{m^a},
\ \ \zeta(\pm a,\pm b,\pm c,\cdots )~=~ \sum^{\infty}_{m=1}
\frac{(\pm 1)^m}{m^a}\, S_{\pm b,\pm c,\cdots}(m-1),  \label{Euler}
\ee

\begin{figure}[t]
\includegraphics[width=0.6\textwidth,height=0.35\textheight,angle=0]{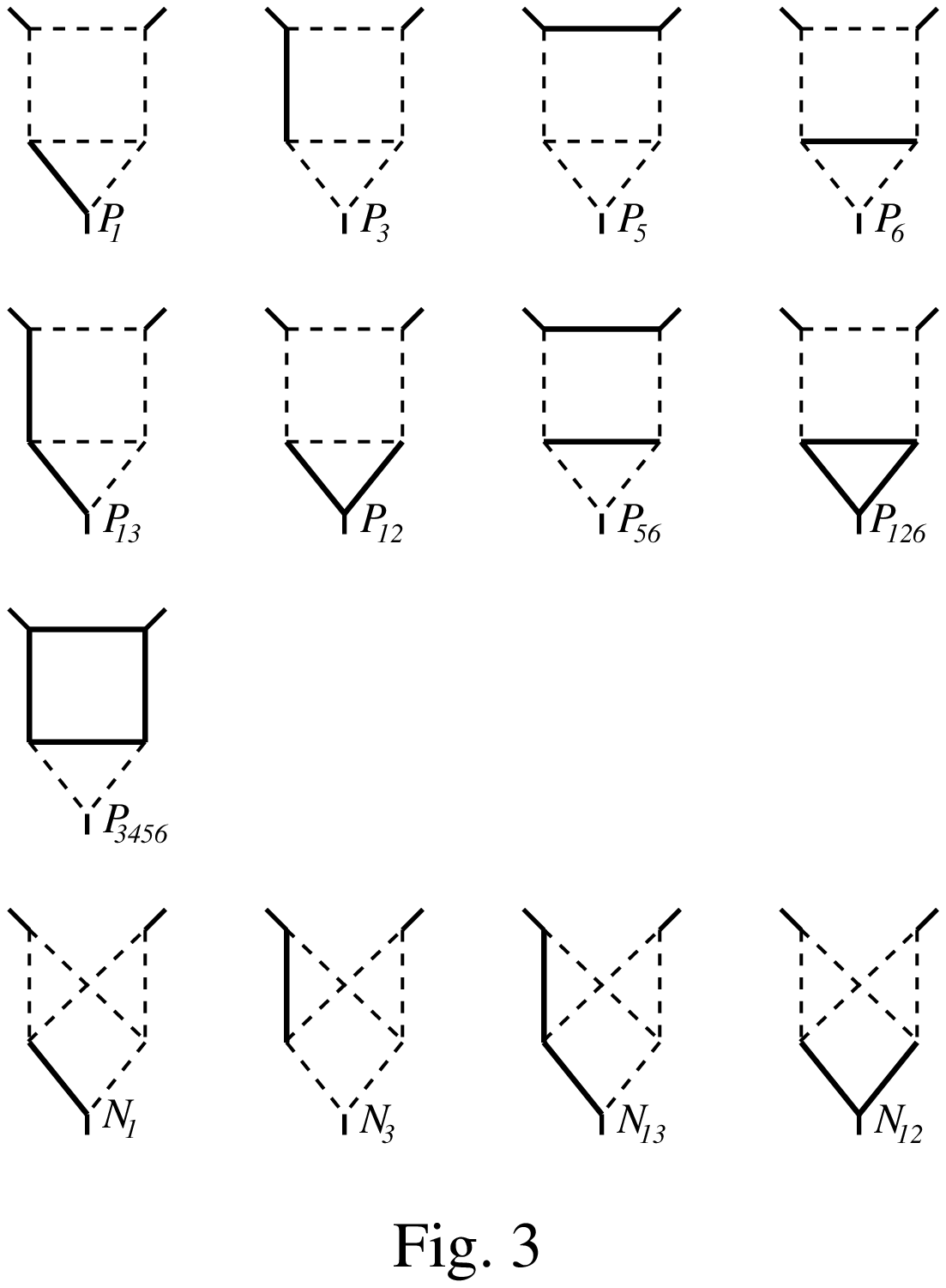}
\end{figure}

For  $2m$-cut
case,
the coefficients $F_{N,k}(n)$ can be more complicated
\begin{eqnarray}
F_{N,k}(n) ~ \sim ~ \frac{S_{\pm a,...}}{n^b},  ~ \frac{V_{a,...}}{n^b}
,  ~ \frac{W_{a,...}}{n^b} \, ,
\label{FI1d}
\end{eqnarray}
where $V_{\pm a,...} \equiv V_{\pm a,...}(j-1)$ and 
$W_{\pm a,...} \equiv W_{\pm a,...}(j-1)$ with
\begin{eqnarray}
V_{a}(j)\ =\ \sum^j_{m=1}
\, \frac{\hat{C}_m}{m^a},
\ \ V_{a,b,c,\cdots}(j)~=~ \sum^j_{m=1}  \,
\frac{\hat{C}_m}{m^a}\, S_{b,c,\cdots}(m),  \label{FI4} \\
W_{a}(j)\ =\ \sum^j_{m=1} \,
\frac{\hat{C}_m^{-1}}{m^a},
\ \ W_{a,b,c,\cdots}(j)~=~ \sum^j_{m=1}  \,
\frac{\hat{C}_m^{-1}}{m^a}\, S_{b,c,\cdots}(m),  \label{FI5}
\end{eqnarray}

The terms $\sim V_{a,...}$ and $\sim W_{a,...}$
can come only in the $2m$-cut
case.
%
The origin of the appearance of these  terms 
is the product of series (\ref{FI1})
with the different 
coefficients $C_n =1$ and
$C_n = \hat{C}_n
$.

As an example, consider two-loop two-point diagrams $I_1$
and $I_{12}$ shown in Fig. 2 and  
studied in \cite{FleKoVe}
\begin{eqnarray}
I_1 &=& \frac{\hat{N}}{q^{2}} \,
\sum_{n=1} \, \frac{x^n}{n} \, \biggl\{
\frac{1}{2} \ln^2 (-x) - \frac{2}{n} \ln (-x)  + \zeta(2)
+2S_2 -2 \frac{S_1}{n} + \frac{3}{n^2} \biggr\} \, ,
\label{FI6a} \\
I_{12} &=& \frac{\hat{N}}{q^{2}} \,
\sum_{n=1} \, \frac{x^n}{n^2} \, \biggl\{\frac{1}{n} +
 \frac{(n!)^2}{(2n)!} \, \biggl(
-2 \ln (-x)  -3 W_1 + \frac{2}{n} \biggr) \biggr\} \, .
\label{FI6c}
\end{eqnarray}

From (\ref{FI6a}) 
one can see that the corresponding functions
$F_{N,k}(n)$ have the form 
\begin{eqnarray}
F_{N,k}(n) ~ \sim ~ \frac{1}{n^{3-N}},~~~~(N\geq 2),
\label{FI8}
\end{eqnarray}
if we introduce the following complexity of the sums ($\Phi=(S,V,W)$) 
\begin{eqnarray}
\Phi_{\pm a} \sim \Phi_{\pm a_1, \pm a_2}
\sim \Phi_{\pm a_1,\pm a_{2},\cdots,\pm a_m}
\sim \zeta_{a} \sim \frac{1}{n^a},~~~~ (\sum_{i=1}^m a_i =a) \, .
\label{FI9}
\end{eqnarray}

The number $3-N$ defines the level of transcendentality (or complexity, or weight)
of the coefficients $F_{N,k}(n)$. The property reduces
strongly the number
of the possible elements in $F_{N,k}(n)$.
The level of transcendentality decreases if we consider the
singular parts of diagrams and/or coefficients in front of
$\zeta$-functions and of
logarithm powers.
Thus, finding the parts we are able to predict the rest, using the ansatz 
based on the results already obtained but containing elements with a higher 
level  of transcendentality.

Other $I$-type integrals in \cite{FleKoVe} have similar form. They have been 
calculated
exactly by DE
method \cite{Kotikov:1990zs,Kotikov:1990kg}.

Now we consider two-loop three-point diagrams, 
$P_5$
and $P_{12}$ shown in Fig. 3 and calculated
in \cite{FleKoVe}:
\begin{eqnarray}
P_5 &=& \frac{\hat{N}}{(q^{2})^2} \,
\sum_{n=1} \, \frac{(-x)^n}{n} \, \biggl\{
-6\zeta_3 + 2(S_1\zeta_2
+6S_3-2S_1S_2+ 4 \frac{S_2}{n}-\frac{S_1^2}{n}
+ 2\frac{S_1}{n^2} \nonumber \\
&&+ \biggl(-4S_2+S_1^2-2\frac{S_1}{n}\biggr)\ln (-x)
+ S_1\ln^2 (-x)\biggl\} \, ,
\label{FI7b} \\
P_{12} &=& \frac{\hat{N}}{q^{2}} \,
\sum_{n=1} \, \frac{x^n}{n^2} \,
 \frac{(n!)^2}{(2n)!} \, \biggl\{
\frac{2}{\ep^2} + \frac{2}{\ep} \biggl(S_1 -3 W_1 + \frac{1}{n} - \ln (-x)
\biggr) +12 W_2 -18 W_{1,1}
\nonumber \\ && -13S_2 + S_1^2- 6S_1W_1 +2 \frac{S_1}{n} +
\frac{2}{n^2}
-2 \bigg(S_1+\frac{1}{n}\biggr)\ln (-x)  + \ln^2 (-x) \biggr\} \, ,
\nonumber
\end{eqnarray}

 Now the coefficients
$F_{N,k}(n)$ have the form
\begin{eqnarray}
  F_{N,k}(n) ~ \sim ~ \frac{1}{n^{4-N}},~~~~(N\geq 3),
\label{FI12}
\end{eqnarray}

The diagram 
$P_5$ 
(and also $P_1$, $P_3$ and $P_6$ in \cite{FleKoVe})
have been calculated
exactly by DE
method \cite{Kotikov:1990zs,Kotikov:1990kg}.
To find the results for  
$P_{12}$ (and also
all others
in \cite{FleKoVe}) we have used the knowledge of the several $n$ terms in the
inverse-mass expansion (\ref{FI1}) (usually less than $n=100$) and
the following arguments: 
\begin{itemize}
\item
If
a two-loop two-point diagram with the
``similar topology'' (for example, 
$I_{12}$ for $P_{12}$ an so on)
has been already calculated, we should consider a
similar set of basic elements for corresponding $F_{N,k}(n)$ of
two-loop three-point diagrams
but with the higher level of complexity.
\item
Let the considered diagram contain singularities and/or powers of logarithms.
Because in front of the leading
singularity, or
the largest power of
logarithm, or the largest $\zeta$-function the coefficients are very simple,
they can be often predicted directly from the first several terms of
expansion.

Moreover, often we can calculate the singular part using another technique
(see \cite{FleKoVe} 
for extraction of $\sim W_1(n)$ part). Then we should expand the
singular parts, find the basic elements and try to use them
(with the corresponding increase of the level of complexity) to predict
the regular part of the diagram. If we have to
find the $\ep$-suppressed terms, we should
increase the level of complexity for the corresponding
basic elements.
\end{itemize}

Later, using the ansatz for $F_{N,k}(n)$ and several terms
(usually, less than 100) in the above
expression, which can be calculated exactly, we obtain the system of
algebraic equations for the parameters of the ansatz. Solving the system, we
can obtain the analytical results for FI without
exact calculations.
To check the results, it is needed only to calculate a few more
terms
in the
above inverse-mass expansion (\ref{FI1}) and compare them with the
predictions of our anzatz with the above fixed coefficients.

So, the considered arguments give a possibility to find the results for many complicated
two-loop three-point diagrams without direct calculations.
Some variations of the procedure have been successfully used for calculating
the Feynman diagrams for many processes (see 
\cite{Fleischer:1997bw,FleKoVe,Kniehl:2005bc,Kniehl:2006bg}).

Note that the properties similar to (\ref{FI8}) and (\ref{FI12}) have been
observed recently \cite{Eden:2012rr} in the so-called double
operator-product-expansion limit of some four-point diagrams.\\

{\bf 3.}~
The coefficients
have the structure   (\ref{FI8}) and (\ref{FI12}) with the rule
(\ref{FI9}). We note that these conditions
reduces strongly the number of
possible harmonic sums. In turn, the restriction
relates with the specific form of the DEs
for the considered FIs. The DEs
formaly can be represented
like \cite{Kotikov:2010gf,Kotikov:2012ac}
\bea
\left((x+a)\frac{d}{dx} - \overline{k}\ep
\right) \, \mbox{ FI } \, = \,
 \mbox{ less complicated diagrams} (\equiv \rm{FI}_1),
\label{Int}
 \eea
with some number $a$ and some function $\overline{k}(x)$.
Such form is generated by IBP procedure for diagrams including
an internal 
$n$-leg one-loop subgraph, in turn containing the product $k^{\mu_1}...k^{\mu_m}$ of its internal momenta $k$ with $m=n-3$.
Indeed, for the usual powers
  $\alpha_i=1+a_i\ep$ with arbitrary $a_i$ of the subgraph propagators, the IBP relation produces the
  coefficient
  \footnote{See, for example, Eqs. (1) and (23) in the first and second papers of Ref. \cite{Kazakov:1986mu}.
    Note that these results were done in the coordinate space and can be applied for the dual diagrams. The
    results in the momentum space are same. Note also that for an internal loop corresponding to the case $n=2$,
it is convenient to take
the index $\alpha_2=2+a_2\ep$.} 
  $D-2\alpha_1 - \sum^p_{i=2} \alpha_i+m \sim \ep$ for $m=n-3$. Important examples of an application of the
  rule are the diagrams in Fig. 2 and the planar ones in Fig. 3 (for the case $n=3$) and the diagrams
  \footnote{
    I thanks Rutger Boels for him information about Ref.  \cite{Gehrmann:2011xn}.}
  in Ref.
  \cite{Gehrmann:2011xn} (for the case $n=3$ and $n=4$).
  However, we note that the results for the nonplanar diagrams in Fig. 3 obey to
  Eq. (\ref{FI12}) but
  their subgraphs
  are not in agreement with the above rule. Perhaps
  the disagreement relates with on-shall vertex of the subgraph but it
  needs additional investigations.

Taking the set of the less complicated Feynman integrals $\rm{FI}_1$ as diagrams having internal $n$-leg subgraphs,
we will have their result stucture similar to above one (\ref{FI12}) but
with the one less level of complexity.

So, the integrals $\rm{FI}_1$ should obey to the following equation
\bea
\left((x+a_1)\frac{d}{dx} - \overline{k}_1\ep
\right) \, \mbox{ FI$_1$ } \, = \,
 \mbox{ less$^2$ complicated diagrams} (\equiv \rm{FI}_2) .
\label{Int.1}
 \eea

 Thus, we will have the set of equations for all Feynman integrals $\rm{FI}_n$ as
 \bea
\left((x+a_n)\frac{d}{dx} - \overline{k}_n\ep
\right) \, \mbox{ FI$_n$ } \, = \,
 \mbox{ less$^{n+1}$ complicated diagrams} (\equiv \rm{FI}_{n+1}),
\label{Int.n}
 \eea
 with the last integral $\rm{FI}_{n+1}$ contains only tadpoles.

 Moreover, following
 to \cite{Henn:2013pwa} we can
recover the above set of the inhomogeneous equations as 
the homogeneous matrix equation
\begin{eqnarray} 
  \frac{d}{dx} \widehat{FI} - \ep  \widehat{K}  \widehat{FI}=0,
\label{Henn}
\end{eqnarray}
 for the vector
 \begin{eqnarray}
   \widehat{FI} =
   \left(  \begin{array}{l}
     \rm{FI}\\  \rm{FI}_1/\ep \\   ... \\  \rm{FI}_n/\ep^n
\end{array} \right)
   \,,
   \nonumber
\end{eqnarray}
 where the matrix $ \widehat{K}$ contains
 the functions $\overline{k}_j/(x+a_j)$ as its elements.
 The form (\ref{Henn}) is very popular now
 (see
 the recent report \cite{Henn:2014qga} and discussion therein)
 
Note that for the real calculations of
$\rm{FI}_{n}$ it is convenient to do the replacement
 \begin{eqnarray}
   \rm{FI}_{n} =  \widetilde{\rm{FI}}_{n}  \overline{\rm{FI}}_{n},
   \nonumber
\end{eqnarray}
 where the term $\overline{\rm{FI}}_{n}$ obeys the corresponding homogeneous equation
 \bea
\left((x+a_n)\frac{d}{dx} - \overline{k}_n\ep
\right) \, \overline{\rm{FI}}_{n}
\, =\, 0,
\label{IntBar.n}
 \eea
 
The replacement simplifies the above equation (\ref{Int.n}) to the following form
\bea
(x+a_n)\frac{d}{dx}  \,  \widetilde{\rm{FI}}_{n} \, = \,   \widetilde{\rm{FI}}_{n+1}
\frac{ \overline{\rm{FI}}_{n+1}}{ \overline{\rm{FI}}_{n}} \, ,
\label{Int.n.si}
 \eea
 having the solution
 \bea
 \widetilde{\rm{FI}}_{n}(x) = \int^x_0 \frac{dx_1}{x_1+a_{n}} \widetilde{\rm{FI}}_{n+1}(x_1)
\frac{ \overline{\rm{FI}}_{n+1}(x_1)}{ \overline{\rm{FI}}_{n}(x_1)}
 \label{Int.n.si1}
 \eea

Usually there are some cancellations in the ratio
$\overline{\rm{FI}}_{n+1}/ \overline{\rm{FI}}_{n}$
and sometimes it is equal to 1. In the last case, the
 equation (\ref{Int.n.si1}) coincides wuth definition of Goncharov Polylogariths (see \cite{Duhr:2014woa} and
 references therein).\\

 The series (\ref{FI6a}), (\ref{FI6c}) and (\ref{FI7b}) can be expressed as combination of the Nilson \cite{Devoto:1983tc}
 and Remiddi-Vermaseren \cite{Remiddi:1999ew}
 polylogarithms with the weight $4-N$ (see \cite{FleKoVe,Fleischer:1997bw}). More complicated cases were considered in
 \cite{Davydychev:2003mv}.


\section{
  ${\mathcal N}=4$ SYM}

The ADs
govern the
Bjorken scaling violation for parton distributions ($\equiv$ matrix elemens of the twist-2 Wilson operators)
in a framework of Quantum Chromodynamics (QCD).

The BFKL
and Dokshitzer-Gribov-Lipatov-Altarelli-Parisi (DGLAP) \cite{DGLAP}
equations resum, respectively, the most important 
contributions
$\sim \alpha_s \ln(1/x_B)$ and $\sim \alpha_s \ln(Q^2/\Lambda^2)$ in different
kinematical regions of the Bjorken variable $x_B$ and the ``mass'' $Q^2$ of the
virtual photon in the lepton-hadron DIS
and, thus, they are the cornerstone in analyses of
the experimental data
from lepton-nucleon and nucleon-nucleon scattering processes.
In the supersymmetric generalization of QCD
the equations are simplified drastically \cite{KL00}.
In the
${\mathcal N}=4$ SYM the eigenvalues of the AD matrix contain only one {\it universal} function
with shifted arguments \cite{LN4,KL}.\\


{\bf 1.} The  three-loop result
\footnote{
Note, that in an accordance with Ref.~\cite{next}
 our normalization of $\gamma (j)$ contains
the extra factor $-1/2$ in comparison with
the standard normalization (see~\cite{KL})
and differs by sign in comparison with one from Ref.~\cite{VMV}.}
for the universal AD
$\gamma_{uni}(j)$
for ${\mathcal N}=4$ SYM is~\cite{KLOV}
\begin{eqnarray}
  \gamma_{uni}(j) ~=~ \hat a \gamma^{(0)}_{uni}(j)+\hat a^2
\gamma^{(1)}_{uni}(j) +\hat a^3 \gamma^{(2)}_{uni}(j) + ... , \qquad \hat a=\frac{\alpha N_c}{4\pi}\,,  \label{uni1}
\end{eqnarray}
where
\begin{eqnarray}
\frac{1}{4} \, \gamma^{(0)}_{uni}(j+2) &=& - S_1,  \label{uni1.1} \\
\frac{1}{8} \, \gamma^{(1)}_{uni}(j+2) &=& \Bigl(S_{3} + \overline S_{-3} \Bigr) -
2\,\overline S_{-2,1} + 2\,S_1\Bigl(S_{2} + \overline S_{-2} \Bigr),  \label{uni1.2} \\
\frac{1}{32} \, \gamma^{(2)}_{uni}(j+2) &=& 2\,\overline S_{-3}\,S_2 -S_5 -
2\,\overline S_{-2}\,S_3 - 3\,\overline S_{-5}  +24\,\overline S_{-2,1,1,1}\nonumber\\
&&\hspace{-1.5cm}+ 6\biggl(\overline S_{-4,1} + \overline S_{-3,2} + \overline S_{-2,3}\biggr)
- 12\biggl(\overline S_{-3,1,1} + \overline S_{-2,1,2} + \overline S_{-2,2,1}\biggr)\nonumber \\
&& \hspace{-1.5cm}  -
\biggl(S_2 + 2\,S_1^2\biggr) \biggl( 3 \,\overline S_{-3} + S_3 - 2\, \overline S_{-2,1}\biggr)
- S_1\biggl(8\,\overline S_{-4} + \overline S_{-2}^2\nonumber \\
&& \hspace{-1.5cm}  + 4\,S_2\,\overline S_{-2} +
2\,S_2^2 + 3\,S_4 - 12\, \overline S_{-3,1} - 10\, \overline S_{-2,2} + 16\, \overline S_{-2,1,1}\biggr)
\label{uni1.5}
\end{eqnarray}
with $S_{\pm a,\pm b, \pm c,...} \equiv  S_{\pm a,\pm b,\pm c,...}(j)$ and 
\begin{eqnarray}
\overline S_{-a,b,c,\cdots}(j) ~=~ (-1)^j \, S_{-a,b,c,...}(j)
+ S_{-a,b,c,\cdots}(\infty) \, \Bigl( 1-(-1)^j \Bigr).  \label{ha3}
\end{eqnarray}

The expression~(\ref{ha3}) is the analytical continuation (to real and 
complex $j$) \cite{AnalCont} of the harmonic sums $S_{-a,b,c,\cdots}(j)$.

The results for $\gamma^{(3)}_{uni}(j)$ \cite{KLRSV,KoReZi}, $\gamma^{(4)}_{uni}(j)$ \cite{LuReVe} and $\gamma^{(5)}_{uni}(j)$
\cite{Marboe:2014sya}
can be obtained from the long-range asymptotic Bethe equations \cite{Staudacher:2004tk} 
for twist-two operators and the additional 
contribution
of the wrapping
corrections.
The similar calculations for the twist-three ADs can be found in \cite{Beccaria:2007pb}.
\\



{\bf 2.} Similary to the eqs. (\ref{FI8})  and  (\ref{FI12})
let us to introduce the
transcendentality level $i$ for the 
harmonic sums $S_{\pm a}(j)$ and
and Euler-Zagier constants $\zeta(\pm a)$
in the following way
\be
S_{\pm a,\pm b,\pm c,\cdots}(j) \sim \zeta(\pm a,\pm b,\pm c,\cdots )
\sim 1/j^i, ~~~~(i=a+b+c+ \cdots)  \label{Tran}
\ee

Then, the basic functions $\gamma
_{uni}^{(0)}(j)$, $\gamma _{uni}^{(1)}(j)$ and $\gamma _{uni}^{(2)}(j)$ are
assumed to be of the types $\sim 1/j^{i}$ with the levels $i=1$, $i=3$ and
$i=5$, respectively. 
A violation of this property could be derived from contributions of 
the terms appearing at a given
order from previous orders of the perturbation theory. Such
contributions could be generated and/or removed by an appropriate finite
renormalization and/or redefinition of the coupling constant. But these terms do not appear in
the ${\overline{\mathrm{DR}}}$-scheme \cite{DRED}.

It is known, that at the first three orders of perturbation theory
(with the SUSY relation for the QCD color factors $C_{F}=C_{A}=N_{c}$) the
most complicated contributions (with $i=1,~3$ and $5$, respectively) are the
same as
in QCD~\cite{VMV}.
This property allows one to find the
universal ADs
$\gamma _{uni}^{(0)}(j)$, $\gamma _{uni}^{(1)}(j)$ and 
$\gamma_{uni}^{(2)}(j)$ without knowing all elements of the AD
matrix~\cite{KL}, which was verified for $\gamma _{uni}^{(1)}(j)$ by the exact 
calculations in~\cite{KoLiVe}.\\

Note that in ${\mathcal N}=4$ SYM the some partial cases of ADs are known also at the large couplings
from
string calculations and AdS/QFT correspondence \cite{Maldacena:1997re}. We would like to note
that if the property of  the maximal transcendentality is existed at low coupling, then sometimes it
appeares at large couplings, too (see, for example, the results for the cusp AD at low \cite{Kotikov:2006ts}
and large  \cite{Basso:2007wd} couplings,
both of which
are based on
the Beisert-Eden-Staudacher equation \cite{Beisert:2006ez}). This is not correct, however, for Pomeron intercept,
which results are lost the property of  the maximal transcendentality at large couplings (see \cite{KLOV,Brower:2006ea,Costa:2012cb}).
The reason of the difference in the results for the cusp AD and Pomeron intercept is not clear now. It needs additional
investigations.


\section{Conclusion}
In the first part of this short review we presented the consideration of Feynman diagramss
(mostly
master integrals), which obey to
the {\it transcendentality principle} (\ref{FI8}),  (\ref{FI9}) and (\ref{FI12}).
Its application leads to the possibility to get the results for most of master integrals
without direct calculations.\\

The second part contains
the universal AD
$\gamma _{uni}(j)$ for
the ${\mathcal N}=4$ SYM in the first three terms of perturbation theory.
All the results have been obtained with using the
{\it transcendentality principle}  (\ref{Tran}).
At the first three orders, the universal ADs
have been extracted directly from the corresponding QCD calculations.
The results for four, five and six loops have been obtained from
the long-range asymptotic Bethe equations \cite{Staudacher:2004tk}
together with some additional terms, so-called {\it wrapping
corrections}, coming in agreement with Luscher approach.
\\

This work was supported by 
RFBR grant 16-02-00790-a.
Author 
thanks Rutger Boels for him information about Ref.  \cite{Gehrmann:2011xn} and
the Organizing Committee of V  International Conference ``Models in Quantum Field Theory''
(MQFT-2015)
for invitation.

\end{document}